\documentclass[12pt]{article}
\newcommand{\dbox}{\,\raise2pt\hbox{\fbox{\rule{2.5pt}{0pt}\rule{0pt}{2.5pt}}}\,}
\newcommand{\qed}{\,\raise0pt\hbox{\mbox{\rule{6.5pt}{6.5pt}}}}
\newcommand{\bra}[1]{\mbox{$\langle #1 |$}}
\newcommand{\ket}[1]{\mbox{$| #1 \rangle$}}
\newcommand{\citen}{\cite}

\newtheorem{proposition}{Proposition}
\begin{document}
\setlength{\baselineskip}{7mm}

\begin{titlepage}
 \begin{normalsize}
  \begin{flushright}
        UT-Komaba/06-12\\
        November 2006
  \end{flushright}
 \end{normalsize}
 \begin{LARGE}
   \vspace{1cm}
   \begin{center}
       New Covariant Gauges in String Field Theory\\
   \end{center}
 \end{LARGE}
  \vspace{5mm}
 \begin{center}
    Masako {\sc Asano}
            \hspace{3mm}and\hspace{3mm}
    Mitsuhiro {\sc Kato}$^{\dagger}$
\\
      \vspace{4mm}
        {\sl Faculty of Liberal Arts and Sciences}\\
        {\sl Osaka Prefecture University}\\
        {\sl Sakai, Osaka 599-8531, Japan}\\
      \vspace{4mm}
        ${}^{\dagger}${\sl Institute of Physics} \\
        {\sl University of Tokyo, Komaba}\\
        {\sl Meguro-ku, Tokyo 153-8902, Japan}\\
      \vspace{1cm}

  ABSTRACT\par
 \end{center}
 \begin{quote}
  \begin{normalsize}
A single-parameter family of covariant gauge fixing conditions in bosonic string field theory is proposed. It is a natural string field counterpart of the covariant gauge in the conventional gauge theory, which includes the Landau gauge as well as the Feynman (Siegel) gauge as special cases. The action in the Landau gauge is largely simplified in such a way that numerous component fields have no derivatives in their kinetic terms and appear in at most quadratic in the vertex.
\end{normalsize}
 \end{quote}

\end{titlepage}
\vfil\eject

\section{Introduction}

Contrary to the ordinary local gauge field theories, covariant string field theory action was first constructed in gauge fixed form by Siegel~\cite{Siegel:1984wx}. After the gauge unfixed action was found~\cite{Siegel:1985tw, Banks:1985ff, Itoh:1985bb} (\citen{Witten:1985cc, Hata:1986zq} for interacting case), the former is recognized as one being fixed with a simple gauge condition $b_0\Phi=0$ called Siegel gauge, where $\Phi$ is a string field and $b_0$ is the zero mode of worldsheet anti-ghost. Twenty years later, we still have no other choice of covariant gauge in a practical sense. Formally we know general method~\cite{Batalin:1984jr} of fixing the gauge symmetry with infinite reducible hierarchy like the one in string field theory. A concrete gauge fixing condition, however, with sufficient simplicity and consistency which Siegel gauge has, is not known to date.

In an intensive study of the tachyon condensation problem, especially in level truncation approach, Siegel gauge has been extensively used and certainly good results have been obtained.\footnote{See Ref.\citen{Taylor:2003gn} for pedagogical review.} In course of these efforts, pathological behavior of this gauge was also encountered. For example, the branch points in the tachyon potential prevent us going beyond a small region of tachyon field value. Therefore an alternative useful gauge choice has been desirable for convincing the known results or extending the analysis.

Also a recent proposal of analytic solution of tachyon vacuum by Schnabl~\cite{Schnabl:2005gv} utilizes the analogue of Siegel gauge in a special conformal frame adapted to the star product of cubic open string field theory~\cite{Witten:1985cc}. There, a modification of the original Siegel gauge is a clue to obtain the closed subset of the space of string field. To further obtain analytic solutions of lower dimensional D-branes or time-dependent ones, more efficient gauge choice may be crucial.

In the present paper, we propose a new covariant gauge in bosonic open string field theory which has simple expression and has, in some aspects, an advantage to the existing one. Actually our new gauge contains a single parameter $a$. Siegel gauge just corresponds to a special value $a=0$. Particularly interesting case is $a=\infty$ which will be mentioned shortly. Varying $a$ we are now able to study a gauge (in-)dependence of various kinds of quantities which were computed previously only in Siegel gauge, so that we can determine which behavior is physical or gauge-artifact.

A new gauge condition we propose is
\begin{equation}
(b_0M + a\,b_0c_0\tilde{Q})\Phi_1=0,
\label{GFa}
\end{equation}
for the ghost number one string field $\Phi_1$, where $a$ is a real parameter, $\tilde{Q}$ and $M$ are defined through the decomposition of the BRST operator $Q$ with respect to the zero mode of worldsheet ghost and anti-ghost\cite{Kato:1983im}:
\begin{equation}
Q = c_0L_0 + b_0M + \tilde{Q}.
\end{equation}
We have to take our gauge parameter $a\ne1$ for free theory, since for $a=1$ the left-hand side of (\ref{GFa}) is invariant under inhomogeneous gauge transformation so that this condition does not fix the gauge for free theory.
A gauge fixing procedure including the gauge condition for the other ghost number sectors (which means ghost, ghost for ghost, $\ldots$) will be described in the later section.

One of the special points of the above family of gauge conditions is $a=0$. Due to the basic property of the operator $M$, (\ref{GFa}) with $a=0$ is turned out to be equivalent to the Siegel gauge $b_0\Phi_1=0$.

Another notable point is $a=\infty$ where the gauge condition reduces to
\begin{equation}
b_0c_0\tilde{Q}\Phi_1=0.
\label{Landau}
\end{equation}
Remarkably this gauge condition (\ref{Landau}) corresponds to the Landau gauge for the massless vector gauge field $A_{\mu}(x)$ contained in the string field, while the Siegel gauge does to the Feynman gauge. So the one parameter extrapolation between them is considered to give general covariant gauge with gauge parameter, say $\alpha$, in gauge field sector which is widely used in the literature.
In order to be definite, let us take free abelian gauge theory. 
The gauge fixed action in consideration is
\begin{equation}
S_{\rm gauge}=\int{d^{26}x}\left[
-\frac{1}{4}F_{\mu\nu}F^{\mu\nu}+B\partial_{\mu}A^{\mu}+\frac{\alpha}{2}B^2
+i\bar{c}\,\partial_{\mu}\partial^{\mu}c
\right]
\end{equation}
where $B$ is the Nakanishi-Lautrup field, $c$ and $\bar{c}$ are the Faddeev-Popov ghost and anti-ghost field respectively. 
Then we can indeed show that the gauge fixing condition (\ref{GFa}) leads to $\alpha=1/(a-1)^2$.

Another characteristic feature of the $a=\infty$ gauge (\ref{Landau}) is that a large number of fields become merely auxiliary fields, {\it i.e.}, kinetic term has no derivative. This can be an advantage in the study of tachyon condensation especially for the space-time dependent solutions. Also these auxiliary fields appear in at most quadratic in the interaction vertex. These facts will more or less simplify the computation compared to the other gauges.

We actually applied new gauges to the level truncation analysis of tachyon condensation and found a remarkably smooth behavior of the tachyon potential around Landau point $a=\infty$ (or sufficiently away from the Siegel point $a=0$), which will be reported in a forthcoming paper\cite{AK2}.

In the following, after preparing the necessary notations and tools in Section 2, we recall the action and the gauge transformation of bosonic open string field theory in Section 3. There we also show that the action can be decomposed into two parts each of which is independently gauge invariant and has quite suggestive form. In Section 4, we propose new gauge conditions including all ghost number sectors, prove the BRST invariance of gauge-fixed action and show a relationship to the covariant gauges in ordinary gauge theory. Finally Section 5 will be devoted to the discussion. Some useful relations as well as technical things are collected in Appendices.

\section{Properties of state space}\label{propss}

In this section, we summarize the properties of the state space 
of open bosonic string theory on $D=26$
as preparation for defining our new gauge fixing conditions.
We first set our notations briefly and then prove some basic facts 
about the property of BRST operator $Q$ (and its constituents $\tilde{Q}$ and $M$) 
and their action on the state space.

The Fock space ${\cal F}(p)$ for bosonic open string is spanned by the states of the form 
\begin{equation}
\ket{f} = \alpha_{-n_1}^{\mu_1}\cdots  \alpha_{-n_i}^{\mu_i}
c_{-l_1}\cdots  c_{-l_{j}}\, b_{-m_1}\cdots  b_{-m_{k}} \,
\ket{0,p;\downarrow}
\label{eq:statephi}
\end{equation}
with $$
0< n_1 \le n_2 \le \cdots \le n_i, \;0< l_1<\cdots < l_{j},
\; 0<m_1<\cdots < m_{k} ,
$$
or the states multiplying $c_0$ on $\ket{f}$ as
\begin{equation}
c_0\ket{f} = \alpha_{-n_1}^{\mu_1}\cdots  \alpha_{-n_i}^{\mu_i}
c_0 c_{-l_1}\cdots  c_{-l_{j}}\, b_{-m_1}\cdots  b_{-m_{k}}
\ket{0,p;\downarrow}.
\label{eq:statexi}
\end{equation}
Here $\alpha_{n}^{\mu}$ is a matter oscillator, $c_n$ and $b_n$ are worldsheet ghost and anti-ghost mode respectively. $\ket{0,p;\downarrow}=\ket{0,p}\otimes\ket{\!\downarrow}$ is annihilated by $\alpha^{\mu}_n$, $c_n$ and $b_n$ ($n>0$).
The Fock ground state $\ket{0,p}$ is the momentum eigenstate and the ghost ground state $\ket{\!\downarrow}$ is related to the SL$(2,R)$ invariant vacuum $\ket{0}$ by $\ket{\!\downarrow}= c_1 \ket{0}$.
We fix the definition of ghost number operator $N^g$ as $N^g=0$ for $\ket{0}$ and
thus $N^g \ket{\!\downarrow} = \ket{\!\downarrow}$.
Then, $N^g \ket{f} = (j-k+1)\ket{f}$ and $N^g c_0\ket{f} = (j-k+2) c_0\ket{f}$ 
for the states (\ref{eq:statephi}) and (\ref{eq:statexi}) respectively.
We define the operator $\tilde{N}^g = \sum_{n> 0}(c_{-n}b_n-b_{-n}c_n) $ 
that counts the ghost number of non-zero frequency modes 
and divide the space ${\cal F}(p)$ by $\tilde{N}^g$ as 
\begin{equation}
{\cal F} = \bigoplus_{\tilde{N}^g = -\infty}^{\infty} 
\left( {\cal F}^{\tilde{N}^g} + c_0 {\cal F}^{\tilde{N}^g}  \right).
\end{equation}
Here ${\cal F}^{\tilde{N}^g}$ (or $c_0{\cal F}^{\tilde{N}^g}$)
consists of states (\ref{eq:statephi}) (or (\ref{eq:statexi})) 
with $j-k = \tilde{N}^g$.
Each space ${\cal F}^{\tilde{N}^g}$ or $c_0{\cal F}^{\tilde{N}^g}$ can further be divided by the level $N$ as ${\cal F}= \oplus_{N\ge 0} {\cal F}_{N}$. 
Level $N$ of the state $\ket{f}$ or $c_0 \ket{f}$ above is   
$ N = n_1+\cdots+ n_i+l_1+ \cdots +l_j +m_1+\cdots +m_k$.

The BRST operator for the theory is given by~\cite{Kato:1983im}
\begin{equation}
Q = \tilde{Q} + c_0 L_0 + b_0 M 
\end{equation}
where 
\begin{equation}
M=-2 \sum_{n>0} n c_{-n} c_{n} 
\end{equation}
and 
\begin{equation}
\tilde{Q}= \sum_{n\ne 0} c_{-n} L_n^{({\rm m})}  
-\frac{1}{2} \sum_{\parbox{12mm}{\tiny\rule{5pt}{0pt}$mn\ne 0$\\$m+n\ne0$}} (m-n)\,:c_{-m} c_{-n} b_{n+m}: .
\end{equation}
Note that $L_0 = L_0^{({\rm m})} + L_0^{({\rm g})}$ gives 
$ L_0 \ket{f} = (\alpha'p^2 + N-1) \ket{f} $.
The operators $M$ and $\tilde{Q}$ act on ${\cal F}^{\tilde{N}^g}$ and map it into ${\cal F}^{\tilde{N}^g+2}$ and ${\cal F}^{\tilde{N}^g+1}$ respectively:
\begin{eqnarray}
M &:&  {\cal F}^{\tilde{N}^g} \rightarrow {\cal F}^{\tilde{N}^g+2} ,
\\
\tilde{Q} &:&  
{\cal F}^{\tilde{N}^g} \rightarrow {\cal F}^{\tilde{N}^g+1}.
\end{eqnarray}
They act on the space $c_0 {\cal F}^{\tilde{N}^g} $ similarly since $[M,c_0]=\{\tilde{Q}, c_0\}=0$.
We also have a relation 
\begin{equation}
\tilde{Q}^2 = -L_0 M
\label{eq:tildeQ2L0M}
\end{equation}
since $Q^2=0$ and $[\tilde{Q}, M] =[\tilde{Q}, L_0]=[L_0, M]=0 $.

The operator $M$ together with 
\begin{equation}
M^-= - \sum_{n>0} \frac{1}{2n}  b_{-n} b_{n} 
\end{equation}
and
\begin{equation}
M_z= \frac{1}{2} \tilde{N}^g = \frac{1}{2} \sum_{n>0} (c_{-n} b_n -b_{-n} c_n)
\end{equation}
constitute SU$(1,1)$ algebra 
\begin{equation}
[M,M^-] = 2 M_z ,\quad [M_z,M] = M, \quad [M_z,M^-] = -M^-.  
\end{equation}

\bigskip

Now we will analyze the properties of state space ${\cal F}^{\tilde{N}^g}$ 
concerning this SU$(1,1)$ algebra and the operator $\tilde{Q}$. 
We can treat the states with or without $c_0$ separately for such an analysis
since the algebra and $\tilde{Q}$ both (anti-)commute with $c_0$. 
In the remaining of this section, we only deal with the space $\oplus_{\tilde{N}^g}{\cal F}^{\tilde{N}^g}$. 

\paragraph{Isomorphism of ${\cal F}^{n}$ between ${\cal F}^{-n}$ and the `inverse'
operator $W_n\quad$}

Every state in the space $\oplus_{\tilde{N}^g}{\cal F}^{\tilde{N}^g}$ can be decomposed into the finite dimensional irreducible representations of the SU$(1,1)$ which are classified by the integer or half-integer spin $s$: 
States belonging to a certain spin $s=k/2$ representation
forms a $(k+1)$-dimensional vector of states such as
\begin{equation}
\ket{f^{(-k)}{} }_{s=\frac{k}{2}}  \, 
\stackrel{M}{\longrightarrow}\,
\ket{f^{(-k+2)}}_{s=\frac{k}{2}} \,
\stackrel{M}{\longrightarrow}\,
\cdots\,
\stackrel{M}{\longrightarrow}\,
\ket{f^{(k-2)}}_{s=\frac{k}{2}} \,
\stackrel{M}{\longrightarrow}\,
\ket{f^{(k)}}_{s=\frac{k}{2}} 
\label{eq:seqrep}
\end{equation}
where $ \ket{f^{(n)}}_{s=\frac{k}{2}} \in  {\cal F}^{n}$ and
$M^- \ket{f^{(-k)}}_{s=\frac{k}{2}}= M \ket{f^{(k)}}_{s=\frac{k}{2}}=0$.
Note also that 
$$(M^-)^j\, M^i \ket{f^{(-k)}}_{s=\frac{k}{2}} \propto M^{i-j} \ket{f^{(-k)}}_{s=\frac{k}{2}}$$
for $1\le i\le k$ and $j\le i$.
The spaces ${\cal F}^{n}$ and ${\cal F}^{-n}$ both consist only of the states belonging to the spin equal or more than $n/2$, {\it i.e.}, 
$s= n/2 +i $ ($i\ge 0$).
From the relation (\ref{eq:seqrep}), each state $ \ket{f^{(-n)}} \in  {\cal F}^{-n}$ reaches a state in ${\cal F}^{n}$ by multiplying $M^n$  and 
conversely, any $ \ket{f^{(n)}} \in  {\cal F}^{n}$ can be written as a form $\ket{f^{(n)}}=M^n \ket{g}$ for some $\ket{g}\in  {\cal F}^{-n}$. 
We also have 
\begin{equation}
M^n \ket{f^{(-n)}} = 0 \;\Rightarrow  \; \ket{f^{(-n)}}=0 .
\label{eq:Mf0f0}
\end{equation}
Thus, ${\cal F}^{-n}$ is isomorphic to ${\cal F}^{n}$ and there exists 
an inverse $W_{n}:{\cal F}^{n}\to {\cal F}^{-n} $ $(n>0)$ which satisfies 
for any $\ket{f^{(-n)} } \in {\cal F}^{-n}$ 
\begin{equation}
W_{n} M^n \ket{f^{(-n)} } =  \ket{f^{(-n)} }. 
\label{eq:WnMn}
\end{equation}
For example, $W_1$ is explicitly written as 
\begin{equation}
W_1 =\sum_{i=0}^{\infty}\, (-1)^{i} 
\left[ \frac{1}{(i+1)!} \right]^2 M^i \left( M^-\right)^{i+1} .
\end{equation}
For $\ket{f^{(n)} } \in {\cal F}^{n}$ ($n>0$), we have
\begin{equation}
W_n \ket{f^{(n)}} = 0 \;\Rightarrow  \; \ket{f^{(n)}}=0,
\label{eq:Wf0f0}
\end{equation}
and 
\begin{equation}
 M^n W_{n} \ket{f^{(n)} } =  \ket{f^{(n)} },
\label{eq:MnWn}
\end{equation}
which will be useful later in gauge fixing procedure.

\paragraph{Properties of $\tilde{Q}$ for $L_0\ne 0$}
We will also use the following properties of $\tilde{Q}$:
\begin{itemize}
\item
For a state $ \ket{f^{(-n)}} \in {\cal F}^{-n}(p)$ $(n>0)$, we have 
\begin{equation}
\tilde{Q} \ket{f^{(-n)}}=0 \;\Rightarrow\; \ket{f^{(-n)}}=0  \qquad \mbox{if}
\quad L_0 \ket{f^{(-n)}} \ne 0 .
\label{eq:Qf0f0}
\end{equation}
\underline{Proof}:~ If $\tilde{Q} \ket{f^{(-n)}} =0$, then $0=\tilde{Q}^2\ket {f^{(-n)}} = -M L_0 \ket{f^{(-n)}}$. 
Thus, $M^{n} \ket{f^{(-n)}}=0 $ and then $\ket{f^{(-n)}}=0 $ 
from (\ref{eq:Mf0f0}).

\item
We also have 
\begin{equation}
 \tilde{Q} \left( {\cal F}^{n}(p) \right)^{L_0\ne 0}
 = \left( {\cal F}^{n+1}(p) \right)^{L_0 \ne 0}
\qquad \mbox{for}\quad n\ge 0.
\label{eq:QFF}
\end{equation}
Here $({\cal F}(p))^{L_0 \ne 0}$ denotes the space $\{\ket{f}\!\in\! {\cal F}(p)\,\big|\,L_0 \ket{f}\!\ne \! 0  \}$. \\
\underline{Proof}:~ The relation $\tilde{Q} {\cal F}^{n} \subset {\cal F}^{n+1}$ is trivial. On the contrary, if $ \ket{f^{(n+1)}}\in {\cal F}^{n+1}$, then $\ket{f^{(n+1)}} = M^{n+1} \ket{g}$ for some $\ket{g} \in {\cal F}^{-(n+1)}$ from the isomorphism of ${\cal F}^{n+1}$ and ${\cal F}^{-(n+1)}$. 
Thus, if $L_0 \ket{f^{(n+1)}} \ne 0$, 
we have $\ket{f^{(n+1)}} = \tilde{Q}\left(-{\tilde{Q}\over L_0 }M^{n}\ket{g}\right)$ for $n\ge 0$. This completes the proof.
\end{itemize}

\section{Gauge invariant action}

In this section, we first remind ourselves of the general form of gauge invariant free action of covariant bosonic string field theory~\cite{Siegel:1985tw,Banks:1985ff,Itoh:1985bb}, and then see 
that the action can be written as a sum of two gauge invariant combinations which will be quite suggestive. 
We also briefly describe the action and the gauge invariance in the interacting case of cubic open string field theory~\cite{Witten:1985cc}.

\subsection{Free action}

Gauge invariant free action of covariant bosonic open string field theory is given by the use of BRST operator $Q$ as~\cite{Siegel:1985tw,Banks:1985ff,Itoh:1985bb} 
\begin{equation}
S^{\rm quad} = -{1\over 2} \langle \Phi_1, Q\Phi_1  \rangle.
\end{equation}
In terms of usual inner product of states, 
$\langle A, B  \rangle = \langle {\rm bpz}(A) | B  \rangle$ 
where ${\rm bpz}(A)$ denotes BPZ conjugation of $A$.
Some properties of inner product and the BPZ conjugation are collected in \ref{app2}.
String field $\Phi_1=\Phi_1 (x^{\mu}(\sigma), c(\sigma), b(\sigma))$ ($0 \le \sigma \le \pi$) 
has ghost number $N^g=1$ (which we explicitly denote as a subscript of $\Phi_1$) and is Grassmann odd. 
It is expanded by Fock space states such that 
\begin{equation}
\Phi_1=\phi^{(0)} +c_0 \omega^{(-1)}
\end{equation}
with 
\begin{eqnarray}
\phi^{(0)} 
&=& \int \frac{d^{26}p}{(2\pi)^{26}} \, \Bigg[ \sum_{ |f\rangle} \ket{f^{(0)}} \,\psi_{|f\rangle }(p)   \Bigg]
\qquad\qquad
\left( \ket{f^{(0)}} \in {\cal F}^{0}(p)\right),
\\
\omega^{(-1)} &=& \int \frac{d^{26}p}{(2\pi)^{26}} \, \Bigg[ \sum_{ |g\rangle} \ket{g^{(-1)}} \,\psi_{|g\rangle}(p)  \Bigg]
\qquad\qquad
\left( \ket{g^{(-1)}} \in  {\cal F}^{-1 }(p)  \right)
\end{eqnarray}
where the coefficient $\psi_{|f\rangle }(p)$ or $\psi_{|g\rangle }(p)$ 
of each state represents the corresponding space-time field.
Hermitian property of the field ($\psi^\dagger = \psi$ or $=-\psi$) is determined 
by that of $S$.
In terms of $\phi^{(0)}$ and $ \omega^{(-1)} $,  
the action is represented as
\begin{equation}
S^{\rm quad} = -{1\over 2} \left( 
\langle \phi^{(0)}, c_0 L_0 \phi^{(0)} \rangle
+2 \langle \tilde{Q} \phi^{(0)} , c_0 \omega^{(-1)} \rangle
+ \langle M \omega^{(-1)} , c_0 \omega^{(-1)} \rangle
\right).
\label{eq:action2}
\end{equation}
It is also written in a simpler form as 
\begin{equation}
S^{\rm quad} = -{1\over 2}
\left\langle  \left( \phi^{(0)} - \frac{1}{L_0}\tilde{Q} \omega^{(-1)}  \right), 
c_0L_0  \left( \phi^{(0)} - \frac{1}{L_0} \tilde{Q} \omega^{(-1)}  \right)  \right\rangle
\end{equation}
if $L_0\ne 0$.

This action is invariant under the gauge transformation
\begin{equation}
\delta\Phi_1 = Q \Lambda_0 .
\label{eq:gaugetr1}
\end{equation}
The gauge parameter $\Lambda_0$ is a Grassmann-even string field of $N^g=0$ 
and is decomposed as 
\begin{equation}
\Lambda_0 = \lambda^{(-1)} + c_0 \rho^{(-2)} 
\end{equation}
where $\lambda^{(-1)}$ and $\rho^{(-2)}$ consist of states in ${\cal F}^{-1}$ and ${\cal F}^{-2}$ respectively.
In terms of $\phi^{(0)}$ and $\omega^{(-1)}$, the transformation is written by
\begin{equation}
\delta \phi^{(0)} = \tilde{Q} \lambda^{(-1)} + M  \rho^{(-2)}
,
\qquad
\delta \omega^{(-1)} = L_0 \lambda^{(-1)} - \tilde{Q} \rho^{(-2)}.
\label{eq:gaugetr2}
\end{equation}
Note that if $L_0\ne 0$, it is also written as
\begin{equation}
\delta(  \phi^{(0)} + c_0 \omega^{(-1)} ) =
\left( \frac{\tilde{Q}}{L_0} +c_0 \right)
(L_0 \lambda^{(-1)} - \tilde{Q} \rho^{(-2)}).
\label{eq:gaugetrL0}
\end{equation}
We see that the combination
\begin{equation}
\zeta^{(1)}=\tilde{Q} \phi^{(0)} +M \omega^{(-1)}
\label{eq:gizeta1}
\end{equation}
is gauge invariant.
Thus from (\ref{eq:WnMn}), we represent $\omega^{(-1)}$ by $\zeta^{(1)}$ and $ \phi^{(0)} $ as
\begin{equation} 
\omega^{(-1)} = W_1(\zeta^{(1)}-\tilde{Q}\phi^{(0)})
\end{equation}
and the action (\ref{eq:action2}) 
is rewritten using $\zeta^{(1)}$ instead of $\omega^{(-1)}$ as
\begin{equation}
S^{\rm quad} = -{1\over 2} \Bigg( 
\langle  \phi^{(0)} , c_0L_0 \phi^{(0)}  \rangle
- \langle \tilde{Q} \phi^{(0)} , c_0 W_1(\tilde{Q}  \phi^{(0)}  )  \rangle
+ \langle \zeta^{(1)} , c_0 W_1 \zeta^{(1)}  \rangle
\Bigg).
\label{eq:action3}
\end{equation}
Remarkably the terms $\langle  \phi^{(0)} , c_0L_0 \phi^{(0)}  \rangle
- \langle \tilde{Q}  \phi^{(0)} , c_0 W_1 (\tilde{Q}  \phi^{(0)}  ) \rangle $
and $\langle \zeta^{(1)} , c_0 W_1 \zeta^{(1)} \rangle$
are both gauge invariant independently. 
The former represents the kinetic terms for each field in $\phi^{(0)}$. The latter is a sum of the terms that are purely quadratic in each field in $\zeta^{(1)}$, which means that $\zeta^{(1)}$ only contains auxiliary fields.

For example, up to level $N=1$, string field is expanded 
by tachyon $\phi$, massless gauge field $A_{\mu}$ and a scalar $\chi$
as\footnote{Here we use real (hermitian) component fields by appropriately inserting `$i$' in the coefficients. See \ref{app3} for the general rule of hermiticity assignment.}
\begin{eqnarray}
\phi^{(0)}_{N\le 1} &=& \int\! \frac{d^{26}p}{(2\pi)^{26}}\,
\frac{1}{\sqrt{\alpha'}}\left( \phi(p) \ket{0,p; \downarrow} + 
A_\mu(p)\alpha_{-1}^\mu \ket{0,p; \downarrow}  \right),
\\
\omega^{(-1)}_{N\le 1} &=& \int\! \frac{d^{26}p}{(2\pi)^{26}} \, 
\frac{i}{\sqrt{2}}\,\chi(p)\, b_{-1} \ket{0,p; \downarrow}  .
\end{eqnarray}
The gauge invariant $\zeta^{(1)}$ is written explicitly as
\begin{equation}
\zeta^{(1)}_{N\le 1} = \int\! \frac{d^{26}p}{(2\pi)^{26}}\, 
\sqrt{2}\left( - i\chi(p) +  A_\mu(p) p^\mu  \right)
c_{-1} \ket{0,p; \downarrow}  
\end{equation}
and the action up to this order becomes a sum of two gauge 
invariant combinations 
\begin{eqnarray}
&& \hspace*{-1.3cm} -\frac{1}{2} \left(
\langle  \phi^{(0)} , c_0L_0 \phi^{(0)}  \rangle
- \langle \tilde{Q}  \phi^{(0)} , c_0 W_1 (\tilde{Q}  \phi^{(0)}  ) \rangle 
\right) \Big|_{N\le 1}
\nonumber\\
&&  =
-\frac{1}{2} \int\! \frac{d^{26}p}{(2\pi)^{26}} \, 
\phi(p)\left( p^2-\frac{1}{\alpha'}\right)\phi(-p) 
-\frac{1}{2} \int\! \frac{d^{26}p}{(2\pi)^{26}} \, 
  A_\mu(p)\left(\eta^{\mu\nu}p^2 - p^\mu p^\nu \right)A_\nu(-p)
\label{eq:kinN1}
\end{eqnarray}
and 
\begin{equation}
-\frac{1}{2} 
\langle \zeta^{(1)} , c_0 W_1 \zeta^{(1)} \rangle
\Big|_{N\le 1}
= 
-\frac{1}{2} \int\frac{d^{26}p}{(2\pi)^{26}} \left( \chi(p) + ip^\mu A_\mu(p) \right)
\left( \chi(-p) + (-i p^\nu) A_\nu(-p) \right).
\end{equation}
Note that the second term of (\ref{eq:kinN1})
exactly coincides with the gauge invariant action of massless vector field $-\frac{1}{4} F_{\mu\nu}F^{\mu\nu}$ where $F_{\mu\nu}=\partial_{\mu}A_{\nu}-\partial_{\nu}A_{\mu}$. 
Gauge transformation up to level $N=1$ is written by gauge parameter $\lambda$ as
\begin{equation}
\delta A_{\mu}(p) =  ip_{\mu}\lambda
,\quad
\delta \chi(p) = p^2 \lambda .
\end{equation}

\subsection{Cubic action}
The action for the cubic open string field theory is given as~\cite{Witten:1985cc}
\begin{equation}
S = -{1 \over 2} \langle \Phi_1, Q \Phi_1  \rangle
-{g \over 3} \langle \Phi_1, \Phi_1 \ast \Phi_1  \rangle 
.
\label{eq:ginvS23}
\end{equation}
Here $\ast$ denotes the star product and $g$ the coupling constant.
This action is invariant under the gauge transformations
\begin{equation} 
\delta \Phi_1 = Q \Lambda_0 + 
g (\Phi_1 \ast \Lambda_0 - \Lambda_0 \ast \Phi_1).
\end{equation}

The action is rewritten in terms of $\phi^{(0)}$ and $\omega^{(-1)}$ as
\begin{eqnarray}
S &=& -{1 \over 2} 
\left( 
\langle  \phi^{(0)} , c_0L_0 \phi^{(0)}  \rangle
- \langle \tilde{Q} \phi^{(0)} , c_0 W_1(\tilde{Q}  \phi^{(0)}  )  \rangle
+ \langle \zeta^{(1)} , c_0 W_1 \zeta^{(1)}  \rangle
\right)
\nonumber\\
& & ~~ -{g \over 3} 
\left(
\langle \phi^{(0)}, \phi^{(0)}\! \ast \phi^{(0)} \rangle 
+ 
3 \langle \phi^{(0)}, \phi^{(0)}\! \ast c_0 \omega^{(-1)} \rangle 
+
3 \langle \phi^{(0)},  c_0 \omega^{(-1)}\! \ast c_0 \omega^{(-1)} \rangle 
\right)
.
\label{eq:cosft}
\end{eqnarray}
Note that the cubic term of $\omega^{(-1)}$ is absent in the action. Thereby, together with the observation on the quadratic term, we can expect that simpler gauge-fixed action can be obtained by gauge fixing more $\phi^{(0)}$ and less $\omega^{(-1)}$.

\section{Gauge fixing}

Now we consider gauge fixing of the action $S$.
The most popular (and in a practical sense essentially only one known and used) gauge condition is the Siegel gauge condition $b_0 \Phi_1=0$ (or $\omega^{(-1)}=0$). It is known that this gauge condition exactly fixes the gauge invariance of the quadratic $g=0$ part of the action if we assume $L_0\ne 0$.

The gauge fixed action for the Siegel gauge is given in the literature as~\cite{Siegel:1984wx, Bochicchio:1986zj, Bochicchio:1986bd, Thorn:1986qj}
\begin{eqnarray}
S_{\rm Siegel} &=& 
-{1 \over 2}  \left\langle \Phi, Q  \Phi \right\rangle
-{g \over 3} \left\langle  \Phi, \Phi \ast \Phi 
\right\rangle 
+ \left\langle b_0 {\cal B} , \Phi  \right\rangle 
\nonumber\\
&=&
-{1 \over 2}  \sum_{n=-\infty}^{\infty} \left\langle \Phi_n, Q  \Phi_{-n+2} \right\rangle
-{g \over 3} \sum_{l+m+n=3} \left\langle  \Phi_l, \Phi_m \ast \Phi_n 
\right\rangle 
+  \sum_{n=-\infty}^{\infty} \left\langle b_0 {\cal B}_{-n+4} , \Phi_{n}  \right\rangle.
\label{eq:Ssiegel}
\end{eqnarray}
Here $\Phi$ and ${\cal B}$ consist of string fields of all ghost numbers as $\Phi= \sum_{n=-\infty}^{\infty} \Phi_n$ and ${\cal B}= \sum_{n=-\infty}^{\infty} {\cal B}_n$.
Note that component fields are taken to be Grassmann odd (even) for $\Phi_{n={\rm even}}$ ($\Phi_{n={\rm odd}}$) so that $\Phi$ is always Grassmann odd.

In principle, this action is obtained from the gauge invariant action (\ref{eq:ginvS23}) by adding Faddeev-Popov ghost terms and gauge fixing terms repeatedly until no gauge symmetry remains in the action.
We instead reach the same action by first extending the action to include the string fields with all ghost numbers as
\begin{equation}
\tilde{S}=
-{1 \over 2}  \left\langle \Phi, Q  \Phi \right\rangle
-{g \over 3} \left\langle  \Phi, \Phi \ast \Phi 
\right\rangle ,
\qquad\qquad \Phi= \sum_{n=-\infty}^{\infty} \Phi_n 
\label{eq:tildeS}
\end{equation}
and then adding gauge fixing term
$\left \langle b_0 {\cal B} , \Phi  \right\rangle $ so as to 
completely fix the gauge invariance of the extended action under the gauge transformation of arbitrary ghost number
\begin{equation}
\tilde{\delta} \Phi = Q \Lambda + 
g (\Phi \ast \Lambda - \Lambda \ast \Phi),
\qquad\qquad \Lambda = \sum_{n=-\infty}^{\infty} \Lambda_n .
\end{equation}

In order to obtain the gauge fixed action for another gauge condition we propose below, we will follow this latter procedure.
That is, for each gauge condition we take,  we will show that we can choose an appropriate operator ${\cal O}$ so that the condition
${\rm bpz}({\cal O})\Phi=0$ 
exactly fixes the gauge invariance of the extended action $\tilde{S}$ for
$g=0$. We then add the term $\langle {\cal O} {\cal B} , \Phi  \rangle$ to $\tilde{S}$ and regard the result as the gauge fixed action. Finally, as a confirmation, we check that the action obtained in this way surely has BRST invariance instead of gauge invariance.

As announced in the introduction, we propose one parameter family of new gauge fixing conditions in this section. We, however, first discuss a special point (Landau-type gauge) of the parameter since there is a technically subtle thing compared to the other points.

\subsection{Landau-type gauge}

\subsubsection{Gauge condition}

A new gauge condition we propose in this subsection is
\begin{equation}
b_0 c_0 \tilde{Q}\Phi_1 =0 \quad (\Leftrightarrow \tilde{Q} \phi^{(0)}=0).
\label{eq:Landaucond}
\end{equation}
For the level $N=1$ gauge field, the condition (\ref{eq:Landaucond}) imposes $p^\mu A_{\mu}(p)=0$.
Thus, this gauge condition is the extension of the Landau gauge for ordinary gauge theory, just as the Siegel gauge is that of the Feynman gauge.

We now show that this new condition exactly fixes the gauge invariance of the quadratic action $S^{\rm quad}$ when $L_0\ne 0$. 

First, any string field $\Phi_1  =\phi^{(0)} + c_0 \omega^{(-1)}$ with $L_0 \ne 0$ can be transformed to satisfy this gauge condition by the gauge transformation (\ref{eq:gaugetr1}) (or (\ref{eq:gaugetr2}) in terms of $\phi^{(0)}$ and $\omega^{(-1)}$) 
with the gauge parameter
\begin{equation}
\Lambda_0 (= \lambda^{(-1)}+c_0 \rho^{(-2)})
 = \frac{1}{L_0} W_1 (\tilde{Q}\phi^{(0)})  
\end{equation}
since
\begin{equation} 
 \tilde{Q}(\phi^{(0)} + \delta \phi^{(0)}) 
=  \tilde{Q} \left(\phi^{(0)} + \tilde{Q} \frac{1}{L_0} W_1 (\tilde{Q}\phi^{(0)})\right) 
=0.
\end{equation}
In the last equation, we have used (\ref{eq:tildeQ2L0M}) and (\ref{eq:MnWn}) for $n=1$.
Furthermore, we can show that there remains no residual gauge transformation within the condition eq.(\ref{eq:Landaucond}): 
If there is such transformation, it is given by the gauge parameter 
$\Lambda_0 = \lambda^{(-1)}+c_0 \rho^{(-2)}$ that satisfies
$\tilde{Q}(\delta_{\Lambda_0} \phi^{(0)})=0 $.
This means that $0=\tilde{Q}(\tilde{Q} \lambda^{(-1)} + M \rho^{(-2)}) 
=M (-L_0 \lambda^{(-1)} +\tilde{Q} \rho^{(-2)} ) $ 
and from (\ref{eq:Mf0f0}), we have $-L_0 \lambda^{(-1)} +\tilde{Q} \rho^{(-2)} =0$. 
If $L_0\ne 0$, this leads to $\delta \Phi_1 = 0$ by (\ref{eq:gaugetrL0}).  

\subsubsection{Gauge fixed action}

Gauge fixed action for this gauge condition is obtained by using the method explained in the beginning of this section.
The resulting action is given as
\begin{eqnarray}
S_{\rm L} &=& 
-{1 \over 2}  \sum_{n=-\infty}^{\infty} \left\langle \Phi_n, Q  \Phi_{-n+2} \right\rangle
-{g \over 3} \sum_{l+m+n=3} \left\langle  \Phi_l, \Phi_m \ast \Phi_n 
\right\rangle 
\nonumber\\
&&
\qquad +  \sum_{n=2}^{\infty} \left(
\left\langle ({\cal O}_{\rm L} {\cal B})_{-n+3} , \Phi_{n}  \right\rangle 
+
\left\langle ({\cal O}_{\rm L} {\cal B})_{n} , \Phi_{-n+3}  \right\rangle 
\right)
\end{eqnarray}
with 
\begin{eqnarray}
({\cal O}_{\rm L} {\cal B})_{n} 
 &=& 
c_0 b_0 M^{n-2} \tilde{Q} {\cal B}_{3-n}
,
\\
({\cal O}_{\rm L} {\cal B})_{-n+3} 
&=&
c_0 b_0 W_{n-1} \tilde{Q} {\cal B}_{n}
+ b_0 {\rm bpz}(1 -{\cal P}_{\tilde{Q}M^{n-2}}) {\cal B'}_{4-n}.
\end{eqnarray}
Here, ${\cal B}_{-n+3}$, ${\cal B}_{n}$ and ${\cal B'}_{4-n}$ ($n>1$) are  Grassmann odd string fields and ${\cal P}_{\tilde{Q}M^{n-2}}$ denotes the projection operator which restricts the fields in ${\cal F}^{n-2}$ ($n>1$) as 
\begin{equation}
{\cal P}_{\tilde{Q}M^{n-2}} \ket{f^{(n-2)}}
\in  \tilde{Q}M^{n-2} {\cal F}^{-n+1}.
\end{equation}
We have 
$(1- {\cal P}_{\tilde{Q}M^{n-2}})  b_0 M^{n-2} \tilde{Q} {\cal B}_{3-n}=0$ and it is followed by 
\begin{equation}
\langle ({\cal O}_{\rm L} {\cal B})_{n} ,
({\cal O}_{\rm L} {\cal B})_{-n+3} \rangle=0.
\end{equation}
From this the action $S_{\rm L}$ can be shown to be invariant under the BRST transformation. We actually show this fact as a following general proposition. An arbitrary operator ${\cal O}$ below is just ${\cal O}_{\rm L}$ for the gauge in consideration. The proposition is sufficiently general so that it will be applied to the wide class of operator ${\cal O}$ including those in the next subsection.
\begin{proposition}
The general gauge fixed action of the form 
\begin{eqnarray}
S_{\rm GF} &=& 
-{1 \over 2}  \sum_{n=-\infty}^{\infty} \left\langle \Phi_n, Q  \Phi_{-n+2} \right\rangle
-{g \over 3} \sum_{l+m+n=3} \left\langle  \Phi_l, \Phi_m \ast \Phi_n 
\right\rangle 
\nonumber\\
&&
\qquad +  \sum_{n=2}^{\infty} \left(
\left\langle ({\cal O} {\cal B})_{-n+3} , \Phi_{n}  \right\rangle 
+
\left\langle ({\cal O} {\cal B})_{n} , \Phi_{-n+3}  \right\rangle 
\right)
\end{eqnarray}
is invariant under the BRST transformation with Grassmann odd parameter $\eta$
\begin{eqnarray}
\delta_B \Phi_n &=& \eta ({\cal O} {\cal B})_{n} \qquad (n>1),
\label{eq:BRS1}
\\
\delta_B \Phi_{n} &=& \eta \Big(Q \Phi_{n-1} + g \sum_{k=-\infty}^{\infty}(\Phi_{n-k} \ast \Phi_{k}) \Big)\qquad (n\le 1),
\label{eq:BRS2}
\\
\quad \delta_B {\cal B}_n &=& 0
\label{eq:BRS3}
\end{eqnarray}
if $\langle ({\cal O} {\cal B})_{n} ,
({\cal O} {\cal B})_{-n+3} \rangle=0$.
\label{proposition1}
\end{proposition}

This is proved straightforwardly by using the relations (\ref{eq:rel1}) and 
(\ref{eq:rel2}) by noting that $\Phi$ is Grassmann odd and ${\cal O}{\cal B}$ is even. \qed

The conditions ${\rm bpz}({\cal O}_{\rm L}) \Phi =0$ given by the action $S_{\rm L}$ restrict each string field $\Phi_{n}(= \phi^{(n-1)} + c_0 \omega^{(n-2)})$ or $\Phi_{3-n}(= \phi^{(2-n)} + c_0 \omega^{(1-n)})$  for $n>1$ as 
\begin{eqnarray}
{\rm bpz} (c_0 b_0 M^{n-2} \tilde{Q}) \Phi_{-n+3}=0 
\quad & (\Leftrightarrow & \tilde{Q} M^{n-2} \phi^{(-n+2)}=0),
\label{eq:condL1}
\\
{\rm bpz} (c_0 b_0 W_{n-1} \tilde{Q}) \Phi_{n}=0
\quad & (\Leftrightarrow & \phi^{(n-1)}=0),
\label{eq:condL2}
\\
 b_0 (1 -{\cal P}_{\tilde{Q}M^{n-2}})  \Phi_{n}=0
\quad & (\Leftrightarrow & \omega^{(n-2)} 
\in \tilde{Q} M^{n-2} {\cal F}^{-n+1}).
\label{eq:condL3}
\end{eqnarray}
We have used (\ref{eq:Wf0f0}) and (\ref{eq:Qf0f0}) to obtain the expression in the parentheses of (\ref{eq:condL2}).
Note also that from the first condition (\ref{eq:condL1}), $\phi^{(-n+2)}$ is restricted to include only spin $(n-2)/2$ states.

The proof that the above conditions exactly fix the gauge symmetry 
$\tilde{\delta} \Phi =Q \Lambda$ of the extended action (\ref{eq:tildeS}) 
for $g=0$ when $L_0\ne 0$ is given in \ref{app1}.

\subsection{One parameter extension}
As we have one parameter family of gauge conditions that interpolate between Feynman gauge and Landau gauge in the gauge theory, we also 
provide intermediate gauge conditions between Feynman-Siegel gauge and the Landau-type gauge for the string field theory.
In this section, we propose such conditions and give the corresponding gauge fixed action.

\subsubsection{Gauge condition}
The intermediate gauge condition we propose here is
\begin{equation}
b_0(M + a c_0 \tilde{Q})\Phi_1 =0 
\quad \left(\Leftrightarrow M \omega^{(-1)}+ a\tilde{Q} \phi^{(0)}=0\right)
\label{eq:conda}
\end{equation}
where $a$ is a real parameter. Though it may not be clear at a glance, the condition is reduced to Siegel gauge for $a=0$ since 
$M \omega^{(-1)}=0$ is equivalent to $\omega^{(-1)} =0$ from (\ref{eq:Mf0f0}).
The Landau-type gauge condition is also obtained by taking $a\to \pm \infty$.
We cannot take $a=1$ as a gauge condition for $S^{\rm quad}$ since the condition takes the form $M \omega^{(-1)} + \tilde{Q} \phi^{(0)}=0$ and the left-hand side just coincides with the gauge invariant combination $\zeta^{(1)}$ of (\ref{eq:gizeta1}).
On the other hand, for each $a\ne 1$, 
by choosing the gauge parameter of the gauge transformation $\delta\Phi_1=Q\Lambda_0$ as
\begin{equation}
\Lambda_0 (= \lambda^{(-1)}+c_0 \rho^{(-2)})
 = \frac{1}{a-1} \frac{1}{L_0} W_1 (M \omega^{(-1)} +a \tilde{Q}\phi^{(0)}),  
\end{equation}
any string field $\Phi_1  =\phi^{(0)} + c_0 \omega^{(-1)}$ with $L_0\ne 0$ is transformed to satisfy the condition (\ref{eq:conda}).
This is shown by using relations (\ref{eq:tildeQ2L0M}) and (\ref{eq:MnWn}).
If there is a gauge transformation $\Lambda_0$ which keeps $\Phi_1$ within the gauge condition, 
\begin{equation}
0=M (\delta\omega^{(-1)}) +a \tilde{Q} (\delta\phi^{(0)})
= M(1-a) (L_0 \lambda^{(-1)} -\tilde{Q} \rho^{(-2)} ) .
\end{equation}
For $a\ne 1$ and $L_0\ne 0$,  
this leads to $L_0 \lambda^{(-1)} -\tilde{Q} \rho^{(-2)} =0$ and $\delta \Phi_1=0$. 
Thus we have shown the validity of the gauge condition for $a\ne 1$.

\subsubsection{Gauge fixed action}
The gauge fixed action for this case should be taken as   
\begin{eqnarray}
S_{a} &=& 
-{1 \over 2}  \sum_{n=-\infty}^{\infty} \left\langle \Phi_n, Q  \Phi_{-n+2} \right\rangle
-{g \over 3} \sum_{l+m+n=3} \left\langle  \Phi_l, \Phi_m \ast \Phi_n 
\right\rangle 
\nonumber\\
&&
\qquad +  \sum_{n=2}^{\infty} \left(
\left\langle ({\cal O}_{a} {\cal B})_{-n+3} , \Phi_{n}  \right\rangle 
+
\left\langle ({\cal O}_{a} {\cal B})_{n} , \Phi_{-n+3}  \right\rangle 
\right)
\end{eqnarray}
where
\begin{eqnarray}
({\cal O}_{a} {\cal B})_{n} 
 &=& 
\left(b_0 M^{n-1}  + a c_0 b_0 M^{n-2} \tilde{Q} \right){\cal B}_{3-n}
,
\\
({\cal O}_{a} {\cal B})_{-n+3} 
&=&
\left( b_0 W_{n-2} + a c_0 b_0 W_{n-1} \tilde{Q} \right) {\cal B}_{n}.
\end{eqnarray}
If $a=0$ is taken, this action is consistently reduced to $S_{\rm Siegel}$ of (\ref{eq:Ssiegel}),
since replacement $b_0 M^{n-1}{\cal B}_{3-n} \to b_0 {\cal B}_{n+1}$ and 
$b_0 W_{n-2}{\cal B}_{n} \to b_0{\cal B}_{-n+4} $ can be performed through the isomorphism by the $M$ and $W_n$ explained in section~\ref{propss}.

We can check that this action is invariant under the BRST transformation (\ref{eq:BRS1})$\sim$(\ref{eq:BRS3}) with ${\cal O}={\cal O}_a$, since for $n>1$ we have 
\begin{eqnarray}
&&\langle ({\cal O}_{a} {\cal B})_{n} , ({\cal O}_{a} {\cal B})_{-n+3} \rangle
\nonumber\\
&& \quad  = 
-  \left\langle {\cal B}_{-n+3} , \,
{\rm bpz}\! \left(b_0 M^{n-1}  + a c_0 b_0 M^{n-2} \tilde{Q} \right) 
\, \left( b_0 W_{n-2} + a c_0 b_0 W_{n-1} \tilde{Q} \right) {\cal B}_{n}
   \right\rangle
\\
&&  \quad = (-)^{n} a  \left\langle {\cal B}_{-n+3} , \,\left( 
b_0 M^{n-1} W_{n-1} \tilde{Q} - b_0 \tilde{Q} M^{n-2} W_{n-2} \right) {\cal B}_{n}
 \right\rangle
\\
&&  \quad = 0.
\end{eqnarray}
In the above equation, we have used (\ref{eq:MnWn}) and 
$$ {\rm bpz}\! \left(b_0 M^{n-1}  + a c_0 b_0 M^{n-2} \tilde{Q} \right) 
= (-)^{n-1} \left(b_0 M^{n-1}  + a b_0 c_0 \tilde{Q}  M^{n-2} \right) .
$$
The condition ${\rm bpz}({\cal O}_a) \Phi_n = 0 $ is explicitly written for each $\Phi_n$ or $\Phi_{-n+3}$ ($n>1$) as
\begin{eqnarray}
b_0 ( M^{n-1}  + a  c_0 \tilde{Q} M^{n-2} ) \Phi_{-n+3}=0 
\quad & \Leftrightarrow & \quad 
M^{n-1} \omega^{(-n+1)} + a \tilde{Q} M^{n-2} \phi^{(-n+2)}=0,
\label{eq:conda1}
\\
b_0 ( W_{n-2}  + a  c_0 \tilde{Q} W_{n-1} )
\Phi_{n}=0
\quad & \Leftrightarrow & \quad 
W_{n-2} \omega^{(n-2)} + a \tilde{Q} W_{n-1} \phi^{(n-1)} =0.
\label{eq:conda2}
\end{eqnarray}
Note that the conditions (\ref{eq:conda1}) and (\ref{eq:conda2}) mean that 
any $\omega$ is written by $\phi$ as
\begin{equation}
\omega^{(-n+1)} = - a W_{n-1} \tilde{Q} M^{n-2} \phi^{(-n+2)},
\qquad
 \omega^{(n-2)} =-  a M^{n-2} \tilde{Q} W_{n-1} \phi^{(n-1)} .
\end{equation}
The validity of the above conditions as a gauge fixing condition for the  extended action (\ref{eq:tildeS}) for $g=0$ is again analyzed in \ref{app1}.

\subsection{Relation to the gauge in ordinary gauge theory}

Let us look at the relation of the above covariant gauge to that in ordinary gauge theory, by taking explicitly the level one fields as follows.
\begin{eqnarray}
\phi^{N=1}&=&\int\frac{d^{26}p}{(2\pi)^{26}}\frac{1}{\sqrt{\alpha'}}\left(
  \gamma(p) b_{-1} +A_{\mu}(p) \alpha^{\mu}_{-1} +i\bar{\gamma}(p) c_{-1}
  \right)\ket{0,p;\downarrow},
\\
\omega^{N=1}&=&\int\frac{d^{26}p}{(2\pi)^{26}}\frac{1}{\sqrt{2}}\left(
  i\chi(p) b_{-1} +u_{\mu}(p) \alpha^{\mu}_{-1} +v(p) c_{-1}
  \right)\ket{0,p;\downarrow}.
\end{eqnarray}
We also expand the field ${\cal B}=c_0{\cal B}_{\omega}$ in the same way as the above:
\begin{eqnarray}
{\cal B}_{\omega}^{N=1}&=&\int\frac{d^{26}p}{(2\pi)^{26}}\frac{1}{\sqrt{2}}\left(
  i\beta_{\chi}(p) b_{-1} +\beta_{u_{\mu}}(p) \alpha^{\mu}_{-1} +\beta_{v}(p) c_{-1}
  \right)\ket{0,p;\downarrow}.
\end{eqnarray}
Then we have $a$-gauge fixed quadratic action for them in coordinate representation as
\begin{eqnarray}
S^{{\rm quad}}_{a,\;N=1}&=&\int{d^{26}x}\left[
-\frac{1}{4}F_{\mu\nu}F^{\mu\nu}-\frac{1}{2}(\chi+\partial_{\mu}A^{\mu})^2
-i\bar{\gamma}\,\partial_{\mu}\partial^{\mu}\gamma-iu_{\mu}\partial^{\mu}\gamma
\right.
\nonumber\\
&&{\rule{40pt}{0pt}}\left.+\beta_{\chi}(\chi+a\partial_{\mu}A^{\mu})
+\frac{1}{2}\beta_{u_{\mu}}(u^{\mu}-a\partial^{\mu}\bar{\gamma})+\frac{1}{4}\beta_vv\right]
\end{eqnarray}
where $F_{\mu\nu}=\partial_{\mu}A_{\nu}-\partial_{\nu}A_{\mu}$.
By use of field redefinitions
$$
\parbox{60mm}{
\begin{eqnarray}
B&=&(a-1)\beta_{\chi}\nonumber\\
\bar{c}&=&(a-1)\bar{\gamma}\nonumber\\
c&=&\gamma\nonumber
\end{eqnarray}
}
\parbox{60mm}{
\begin{eqnarray}
\tilde{\chi}&=&\chi+\partial_{\mu}A^{\mu}-\beta_{\chi}\nonumber\\
\tilde{u}^{\mu}&=&u^{\mu}-a\partial^{\mu}\bar{\gamma}\nonumber\\
\tilde{\beta}_{u_{\mu}}&=&\beta_{u_{\mu}}+2i\partial_{\mu}\gamma,\nonumber
\end{eqnarray}
}
$$
the above action can be written into the well-known form plus decoupled auxiliary fields' term
\begin{equation}
S^{{\rm quad}}_{a(\alpha),\;N=1}=\int{d^{26}x}\left[
-\frac{1}{4}F_{\mu\nu}F^{\mu\nu}+B\partial_{\mu}A^{\mu}+\frac{\alpha}{2}B^2
+i\bar{c}\,\partial_{\mu}\partial^{\mu}c
\right.
\left.-\frac{1}{2}\tilde{\chi}^2
+\frac{1}{2}\tilde\beta_{u_{\mu}}\tilde{u}^{\mu}+\frac{1}{4}\beta_vv
\right].
\end{equation}
First four terms are the usual abelian gauge theory action in covariant gauge with the gauge parameter $\alpha$ which is related to $a$ as
\begin{equation}
\alpha=\frac{1}{(a-1)^2}.\label{eq:alpha}
\end{equation}
Here $B$ is the Nakanishi-Lautrup field, $c$ and $\bar{c}$ are the Faddeev-Popov ghost and anti-ghost field respectively. We can read off from these that $a=0$ ($\alpha=1$) corresponds to Feynman gauge and $a=\infty$ ($\alpha=0$) does to Landau gauge. Also we recognize that $a=1$ ($\alpha=\infty$) is the point where the gauge is not fixed for free theory.

For the Landau-type gauge, we can obtain the level one part of the action similarly for the above $a$-gauge. In this case, however, we need ${\cal B}_2'=c_0{\cal B}_{\omega}'$ field in addition to the above ${\cal B}^{N=1}$. 
By noting that the projection operator ${\cal P}^{N=1}_{\tilde{Q}}$ on ${\cal B}_{\omega}'$ is written as 
$$
{\rm bpz} (1-{\cal P}^{N=1}_{\tilde{Q}})   {\cal B}_{\omega}' 
= \left(1- \frac{L_1 L_{-1}}{2 L_0} \right) {\cal B}_{\omega}' ,
$$
we obtain the explicit form of $S_{{\rm L},\; N=1}^{\rm quad}$,
which coincides with $S^{\rm quad}_{a(\alpha),\;N=1}$ for $\alpha=0$ ($a\to \infty$) 
after performing appropriate field redefinitions.

Note that the relation (\ref{eq:alpha}) is valid for free theory. For interacting case, the form of action does not directly coincide between the string field level and the effective field theory level because the latter is obtained by integrating out the higher mass level fields. 
In fact, free action in general has a symmetry $S^{\rm quad}_{a}=S^{\rm quad}_{2-a}$, though the interacting action does not. 
In addition the form of gauge transformation is also different from free case due to homogeneous term, so that the gauge non-fixed point will be shifted depending on the value of the fields. This is indeed seen in the level truncated scalar effective potential~\cite{AK2}.

\subsection{Properties of the gauge conditions}

To clarify the properties of each gauge condition we have defined, we especially choose the distinctive two types of condition, the Siegel gauge ($a=0$) and the Landau-type gauge ($a\to \infty$), and compare the properties of them. 
Other gauge condition given by finite $a$ has similar property with the Siegel gauge. 

For the Siegel gauge, all the component fields in $\phi^{(0)}$ of $\Phi_1=\phi^{(0)}+c_0\omega^{(-1)}$ remain in the gauge fixed action, 
while $\omega^{(-1)}$ is completely gauged away. 
In the action $S_{a=0}$, each field $\psi_{\phi}(p)$ in $\phi^{(0)}$ equally has kinetic term through $-\frac{1}{2} \langle \phi^{(0)}, c_0 L_0 \phi^{(0)} \rangle$. 

For the Landau gauge, on the other hand, all the fields in $\omega^{(-1)}$ and the part of $\phi^{(0)}$ field ($\phi^{'(0)}$) satisfying $\tilde{Q} \phi^{'(0)}=0$ for $L_0\ne 0$ remain in the gauge fixed action%
\footnote{We can naturally check that the degrees of freedom for of $\phi^{(0)}$ and $(\phi^{'(0)},\omega^{(-1)})$ are the same since for $L_0\ne 0$ we have a relation 
$
{\cal F}^{0} = \tilde{Q} {\cal F}^{-1} \oplus ({\cal F}^{0})^{\tilde{Q}} 
$
from (\ref{eq:QFF}).
(Here  $({\cal F}^{0})^{\tilde{Q}} =\{\ket{f^{(0)}} \,|\, \tilde{Q}\ket{f^{(0)}}=0 \}$.) 
}.
The $\Phi_1$ part of quadratic action for this gauge becomes
$ 
-\frac{1}{2}( 
\langle \phi^{'(0)}, c_0 L_0 \phi^{'(0)} \rangle +
\langle M \omega^{(-1)} , c_0 \omega^{(-1)} \rangle)
$, which means that all fields $\psi_{\omega}(p)$ in $\omega^{(-1)}$ are auxiliary fields since there is no derivative for each $\psi_{\omega}(p)$ in the action.
In addition, we see from (\ref{eq:cosft}) that there appear 
at most quadratic terms of $\omega^{(-1)}$ even in the full action.
Thus, by choosing the Landau-type gauge condition, the number of dynamical 
fields is extremely reduced compared to the other gauges 
and the analysis of the theory would become simpler.

\section{Discussions}
We have proposed one parameter family of covariant gauge conditions  
for open bosonic string field theory that includes Siegel gauge as a special case ($a=0$).
For the massless gauge field $A_\mu$ contained in the string field $\Phi_1$,  
these gauge conditions precisely correspond to a family of conventional covariant gauges parameterized by $\alpha$, as 
the Siegel gauge does to the Feynman gauge.

In the family of gauge conditions, the $a = \infty$ gauge (which corresponds to Landau gauge in gauge theory) 
especially has several striking features as we have mentioned in previous section.
The features come from the fact that the $\omega^{(-1)}$ part of string field $\Phi_1$ has no derivatives in 
the free action and also it appears in at most quadratic in the interaction terms,
since the $a= \infty$ gauge condition only restricts the $\phi^{(0)}$ part of $\Phi_1$ and leaves whole $\omega^{(-1)}$ untouched.
(The Siegel gauge is exactly in a opposite position since it completely eliminates $\omega^{(-1)}$ and leaves whole  
$\phi^{(0)}$ untouched.)
In the ordinary gauge theory, we know that gauge parameter does not receive renormalization under the Landau gauge. 
It might be possible to find a similar property for the 
$a=\infty$ gauge counterpart of string field theory.
From these advantages of this gauge, we may expect that the quantum analysis 
becomes much simpler than in other gauge.

The form of the quadratic action of $a=\infty$ gauge is reminiscent of the action for the string field based on the old covariant quantization~\cite{Banks:1985ff}:
\begin{equation}
S = {1\over 2}\Phi(L_0-1)\Phi, \qquad 
\left(\mbox{with}\quad L_n\Phi = 0, \quad n\ge 1\right)
\end{equation}
since the condition $\tilde{Q}\phi^{(0)}=0$ reduces to 
$L_{n}\phi^{(0)}=0$ $(n\ge 1)$ if there is no ghost fields ($c_{-n}$ or $b_{-n}$) in $\phi^{(0)}$. 
It would be possible to take explicit basis of the set of states satisfying the condition $\tilde{Q} \ket{f}=0$ 
as well as we have analyzed in the old covariant theory for the states satisfying $L_n\ket{f}=0$ ($n\ge 1$)~\cite{Asano:2005fm}. 
With such a basis of states, analysis of the $a=\infty$ gauge 
would become easier in various situations.

As an application of our new gauge conditions, 
we can analyze the problem of tachyon condensation.
The problem has been mostly analyzed in the Siegel gauge so far by using level truncation method and various 
significant results have been obtained, whereas limitations of such an analysis have also been recognized.
Our one parameter family of new gauge conditions makes it possible to
analyze the problem especially from the viewpoint of gauge dependence (or independence). 
In fact, we analyze properties of tachyon potential under various gauges 
using the level truncation method and find out some interesting results which will be reported in a separate paper~\cite{AK2}.

We have also presented a rather general method to obtain the proper gauge fixed action for 
given gauge fixing condition.
By using the technique, we would be able to further extend our analysis to more general (covariant or non-covariant) 
gauge fixing conditions.
For example, consider the condition ${\cal B}_0\Phi_1=0$ that is used for giving an analytic solution of tachyon 
condensation in~Ref.\citen{Schnabl:2005gv}.
Here ${\cal B}_0$ is the zero mode of the worldsheet $b(\tilde{z})$ field in the 
coordinate $\tilde{z}$ taken as different from canonical one.
Thus the condition is the counterpart of the Siegel gauge in this conformal frame.
It will be possible to find exact gauge fixed action for this gauge condition, or to further extend the condition to 
one parameter family.
Such an analysis may provide some new insights on the problem related to tachyon condensations. 

We have only analyzed open string field theory in the present paper.
The extension to closed string field theory or superstring field theory of our new gauge conditions is left in the future.

\section*{Acknowledgements}
We would like to thank H.~Hata and T.~Takahashi for useful comments.
The work is supported in part by the Grants-in-Aid for Scientific Research (17740142~[M.A.], 13135205 and 16340067~[M.K.]) from the Ministry of Education, Culture, Sports, Science and Technology (MEXT) and from the Japan Society for the Promotion of Science (JSPS).
\appendix 
\renewcommand{\theequation}{\Alph{section}.\arabic{equation}}
\def\thesection{Appendix~\Alph{section}}
\setcounter{equation}{0}
\def\thesection{Appendix~\Alph{section}}
\section{Gauge fixing condition ${\rm bpz}({\cal O})\Phi=0 $ for the extended action $\tilde{S}$}
\label{app1}
We will show that the conditions ${\rm bpz}({\cal O}_{\rm L})\Phi=0 $ [$\Leftrightarrow$ (\ref{eq:condL1}) $\sim$ (\ref{eq:condL3})] and 
${\rm bpz}({\cal O}_a) \Phi=0 $ [$\Leftrightarrow$ (\ref{eq:conda1}) and (\ref{eq:conda2})] are both valid for fixing the gauge symmetry $\tilde{\delta}\Phi= Q \Lambda$ of the quadratic ($g=0$) part of the action $\tilde{S}$ if we assume $L_0 \ne 0$.
We first take the condition ${\rm bpz}({\cal O}_a) \Phi=0 $ and confirm that 
(a) any $\Phi$ can be transformed to satisfy ${\rm bpz}({\cal O}_a)\Phi=0 $ 
and that (b) there is no residual gauge symmetry within the condition.
Then we show the same statements for the condition ${\rm bpz}({\cal O}_{\rm L})\Phi=0 $ by recognizing the condition as $a\to \infty$ limit of ${\rm bpz}({\cal O}_a) \Phi=0 $. 
\paragraph{The condition ${\rm bpz}({\cal O}_a)\Phi=0$ ($a\ne 1$)}
\begin{description}
\item[(a)]
For $n\le 1$, we can show that any $\Phi_n$ is transformed to satisfy the condition (\ref{eq:conda1}) by the extended gauge transformation
$\tilde{\delta}\Phi_n= Q \Lambda_{n-1}$ 
with gauge parameters
\begin{equation}
\Lambda_{n-1} =
 \frac{1}{a-1}\frac{1}{L_0} W_{-n+2} 
\left(M^{-n+2}b_0 +  a\tilde{Q} M^{-n+1} b_0c_0 \right) \Phi_{n} .
\label{eq:aLambda1}
\end{equation}

To analyze the $n>1$ case, 
we first divide ${\cal F}^{n}$ for $n\ge 0$ as
\begin{equation}
{\cal F}^{n} 
= \left( {\cal F}^{n}\right)^{\tilde{Q}}
\oplus {\cal G}^{n}
\label{eq:Fdivp}
\end{equation}
where $\left( {\cal F}^{n}\right)^{\tilde{Q}}$ consists of states $f^{(n)}\in {\cal F}^{n}$ satisfying $\tilde{Q}f^{(n)}=0$ and ${\cal G}^{n}$ is the complement of
$\left( {\cal F}^{n}\right)^{\tilde{Q}}$. 
Also we can divide the space ${\cal F}^{-n}$ for $n\ge 0$
into $\tilde{Q} {\cal F}^{-n-1} $ and its complement ${\cal K}^{-n}$
as 
\begin{equation}
{\cal F}^{-n} 
= \left( \tilde{Q}{\cal F}^{-n-1}\right)
\oplus {\cal K}^{-n}.
\end{equation}
If we limit ourselves to the states with $L_0 \ne 0$,
there is an isomorphism 
$(\tilde{Q}{\cal F}^{-n-1})^{L_0\ne 0} \sim ({\cal G}^n )^{L_0\ne 0}$ 
and thus 
$({\cal K}^{-n})^{L_0\ne 0} \sim (({\cal F}^n)^{\tilde{Q}} )^{L_0\ne 0}$. 
The former is shown from 
$M^{n+1} {\cal F}^{-n-1} ={\cal F}^{n+1}$ and 
$({\cal F}^{n+1})^{L_0\ne 0} = ( \tilde{Q} {\cal F}^n )^{L_0\ne 0}
 = ( \tilde{Q} {\cal G}^n )^{L_0\ne 0}$.
Explicit relations between these spaces are given as   
\begin{equation}
 ( M^{n} \tilde{Q} {\cal F}^{-n-1})^{L_0\ne 0}   = ({\cal G}^n )^{L_0\ne 0},
\qquad  W_n  (({\cal F}^n)^{\tilde{Q}} )^{L_0\ne 0} =({\cal K}^{-n})^{L_0\ne 0} .
\label{eq:isoadd}
\end{equation}

From the above discussion, for $\Phi_n = \phi^{(n-1)} + c_0 \omega^{(n-2)}$ with $n>1$, we can rewrite $\omega^{(n-2)}$ as 
\begin{equation}
\omega^{(n-2)}= M^{n-2}\tilde{Q} f^{(-n+1)} + \tilde{\omega}^{(n-2)},
\quad \left( f^{(-n+1)} \in {\cal F}^{-n+1}, \; 
\tilde{\omega}^{(n-2)} \in ({\cal F}^{n-2})^{\tilde{Q}}  \right)
\end{equation}
according to (\ref{eq:Fdivp}) and (\ref{eq:isoadd}).   
Then, we see that any $\Phi_n$ is transformed to satisfy the condition 
(\ref{eq:conda2}) by the gauge transformation if we choose the gauge parameter as
\begin{equation}
\Lambda_{n-1} =  \frac{1}{a-1}\frac{1}{L_0} M^{n-2}\tilde{Q} 
\left(f^{(-n+1)} + a W_{n-1} \phi^{(n-1)} \right)
-\frac{1}{L_0}\tilde{\omega}^{(n-2)}
.
\label{eq:aLambda2}
\end{equation}
\item[(b)]
For $\Phi_n$ ($n\le 1$), the gauge symmetry under the condition (\ref{eq:conda1}) is represented by the gauge parameter $\Lambda_{n-1}=\lambda^{(n-2)}+c_0 \rho^{(n-3)}$ satisfying
$(1-a)M^{2-n}(L_0 \lambda^{(n-2)} - \tilde{Q} \rho^{(n-3)})=0$ if any.
This reduces to $ L_0 \lambda^{(n-2)} -\tilde{Q} \rho^{(n-3)}=0$ from (\ref{eq:Mf0f0}) and thus $Q \Lambda_{n-1}=({\tilde{Q}\over L_0}+ c_0)(L_0 \lambda^{(n-2)} -\tilde{Q} \rho^{(n-3)}) =0$, which means that there is no residual symmetry.

For $\Phi_n$ ($n>1$), the gauge symmetry under the condition (\ref{eq:conda2}) is represented by $\Lambda_{n-1}$ with 
\begin{equation}
\left(W_{n-2}+a \frac{1}{L_0}\tilde{Q}W_{n-1}\tilde{Q} \right)  (L_0 \lambda^{(n-2)} -\tilde{Q} \rho^{(n-3)} )=0. 
\label{eq:condab}
\end{equation}
We represent $L_0 \lambda^{(n-2)} -\tilde{Q} \rho^{(n-3)}\in {\cal F}^{n-2}$ by $h^{(-n+1)}\in {\cal F}^{-n+1}$ and $\tilde{\theta}^{(n-2)}\in ({\cal F}^{n-2})^{\tilde{Q}}$ as 
$$ L_0 \lambda^{(n-2)} -\tilde{Q} \rho^{(n-3)} = 
M^{n-2}\tilde{Q} h^{(-n+1)} +\tilde{\theta}^{(n-2)} .
$$
Then (\ref{eq:condab}) is rewritten as 
$$
(1-a)\tilde{Q} h^{(-n+1)} + W_{n-2} \tilde{\theta}^{(n-2)}=0.
$$
When $a\ne 1$, this leads to $h^{(-n+1)}=\tilde{\theta}^{(n-2)}=0$.
Thus we conclude that there remains no residual symmetry.
\end{description}

\paragraph{The condition ${\rm bpz}({\cal O}_{\rm L})\Phi=0$
($a\to \infty$ limit of  ${\rm bpz}({\cal O}_{a})\Phi=0$)}
\begin{description}
\item[(a)]
We can show that any $\Phi_n$ is transformed to satisfy the conditions (\ref{eq:condL1}) and  (\ref{eq:condL2}) by $\tilde{\delta}\Phi_n= Q \Lambda_{n-1}$ with 
\begin{equation}
\Lambda_{n-1} =
\left\{
\begin{array}{ll}
\displaystyle
 \frac{1}{L_0} W_{-n+2} \tilde{Q} M^{-n+1} \phi^{(n-1)} & (n\le 1)
\\[10pt]
\displaystyle
 \frac{1}{L_0} M^{n-2}\tilde{Q} W_{n-1} \phi^{(n-1)} 
-\frac{1}{L_0}\tilde{\omega}^{(n-2)}
& (n > 1)
\label{eq:L12}
\end{array}
\right. .
\end{equation}
Note that the above $\Lambda_{n-1}$ coincides with 
$\lim_{a\to \infty}\Lambda_{n-1}$ of (\ref{eq:aLambda1}) and (\ref{eq:aLambda2}) respectively for $n\le 1$ and $n>1$.

\item[(b)]
We can prove the absence of residual gauge symmetry 
just by considering $a\to \infty$ limit of the analysis for $|a| < \infty$ case.
\end{description}


\def\thesection{Appendix~\Alph{section}}
\section{Useful formulas}  \label{app2}
\setcounter{equation}{0}
BPZ conjugation for states is defined by the following equations:
$$ {\rm bpz}(\ket{0,p}) = \bra{0,-p},
\quad
{\rm bpz}(\ket{0}) = \bra{0},
$$ 
$$
{\rm bpz}(b_{-n}) = (-1)^n b_{n},\quad
{\rm bpz}(c_{-n}) = (-1)^{n-1} c_{n},\quad
{\rm bpz}(\alpha^{\mu}_{-n}) = (-1)^{n-1} \alpha^{\mu}_{n},
$$
$$
{\rm bpz}\,(\alpha \beta) = (-1)^{|\alpha| |\beta|}  {\rm bpz}(\beta) \, {\rm bpz}(\alpha) 
$$
where $|\alpha|$ and $|\beta|$ are the Grassmann parity of $\alpha$ and $\beta$ respectively, 
and $ |\alpha| =1$ ($ |\alpha|=0 $) for Grassmann odd (even) operator $\alpha$.
Note that $\ket{0}$ is Grassmann even.
The \hbox{(anti-)}commutation relations among $\alpha_{n}^\mu$, $c_{n}$ and $b_{n}$ are given as 
$$[\alpha_{m}^\mu, \alpha_{n}^\nu ] =m\eta^{\mu\nu}\delta_{m+n,0},
\qquad \{b_{m},c_n \}= \delta_{m+n,0},
\quad \{b_m,b_n\}=\{c_m,c_n\}=0.$$
Normalization of states we use is 
$$\bra{0, p;\downarrow} c_0 \ket{0,p';\downarrow }=(2\pi)^{26} \delta(p-p'). $$
For string fields $A$ and $B$, there are relations
\begin{equation}
Q (A \ast B) = (Q A) \ast B + (-1)^A A \ast (Q B)
\label{eq:rel1}
\end{equation}
and
\begin{equation}
\langle A, B  \rangle = (-1)^{|A||B|} \langle  B, A  \rangle ,
\label{eq:rel2}
\end{equation}
which we use for the proof of gauge or BRST invariance of the action.

\section{Hermiticity of fields} \label{app3}
\setcounter{equation}{0}

Here we note how to assign hermiticity of each field in the string field.
Any string field $\Phi$ is expanded in terms of the first-quantized states $\ket{s(p)}$ as
\begin{equation}
\Phi = \int\frac{d^{26}p}{(2\pi)^{26}}\sum_s\ket{s(p)}\psi^s(p)
\end{equation}
where $\psi^s(p)$ is a component field in momentum representation. 
Using the fact that the BPZ conjugation of the $\ket{s(p)}$ is related to the hermitian conjugation of $\ket{s(-p)}$ by a sign factor $\varepsilon_s$, {\it i.e.}, $\bra{\mbox{bpz}(s(p))}\equiv\mbox{bpz}(\ket{s(p)})=\varepsilon_s\bra{s(-p)}$, let us consider
\begin{equation}
\langle\Phi, {\cal H}\Phi\rangle =
 \int\frac{d^{26}p}{(2\pi)^{26}}\frac{d^{26}p'}{(2\pi)^{26}}\sum_{s,s'}
 \psi^{s'}(p')\bra{\mbox{bpz}(s'(p'))}{\cal H}\ket{s(p)}\psi^{s}(p)
\end{equation} 
Then for any hermitian operator ${\cal H}(={\cal H}^{\dagger})$
diagonal in $p^{\mu}$, we have
\begin{eqnarray}
\langle\Phi, {\cal H}\Phi\rangle &=&
 \int\frac{d^{26}p}{(2\pi)^{26}}\frac{d^{26}p'}{(2\pi)^{26}}\sum_{s,s'}
 \psi^{s'}(p')\varepsilon_{s'}\bra{s'(-p')}{\cal H}(p)\ket{s(p)}
 \psi^{s}(p)\\
&=&
 \int\frac{d^{26}p}{(2\pi)^{26}}\sum_{s,s'}
 \psi^{s'}(-p)\varepsilon_{s'}\bra{s'(p)}{\cal H}(p)\ket{s(p)}
 \psi^s(p)\\
\langle\Phi, {\cal H}\Phi\rangle^* &=&
 \int\frac{d^{26}p}{(2\pi)^{26}}\sum_{s,s'}
 (\psi^{s}(p))^*\left(\varepsilon_{s'}\bra{s'(p)}{\cal H}(p)\ket{s(p)}\right)^*
 (\psi^{s'}(-p))^*\\
&=&
\int\frac{d^{26}p}{(2\pi)^{26}}\sum_{s,s'}
 (\psi^{s}(p))^*\bra{s(p)}{\cal H}(p)\ket{s'(p)}\varepsilon_{s'}
 (\psi^{s'}(-p))^*\\
&=&
\int\frac{d^{26}p}{(2\pi)^{26}}\sum_{s,s'}
 (\psi^{s}(p))^*\varepsilon_{s}\bra{\mbox{bpz}(s(-p))}{\cal H}(p)\ket{s'(p)}
 \varepsilon_{s'}(\psi^{s'}(-p))^*
\end{eqnarray} 
Thus $\langle\Phi, {\cal H}\Phi\rangle^* = \langle\Phi, {\cal H}\Phi\rangle$ if we assign $(\psi^s(p))^* = \varepsilon_s\psi^s(-p)$ or equivalently $(\psi^s(x))^* = \varepsilon_s\psi^s(x)$ in the coordinate representation. Also this assignment is equivalent to the relation $\mbox{bpz}(\Phi)=\Phi^*$. This can be immediately applied to the kinetic term of the string field by taking $Q$ as ${\cal H}$.  

The sign factor $\varepsilon_s$ for a state
\begin{equation}
\ket{s(p)} = \alpha_{-n_1}^{\mu_1}\cdots  \alpha_{-n_i}^{\mu_i}
c_{-l_1}\cdots  c_{-l_{j}}\, b_{-m_1}\cdots  b_{-m_{k}} \,
\ket{0,p;\downarrow}
\label{eq:stateS}
\end{equation}
with $$
0< n_1 \le n_2 \le \cdots \le n_i, \;0\le l_1<\cdots < l_{j},
\; 0<m_1<\cdots < m_{k}
$$
is explicitly computed as
\begin{equation}
\varepsilon_s=(-)^{N+i-\frac{j-k}{2}+\frac{(j+k)^2}{2}}
\end{equation}
where $N=\sum_an_a+\sum_bl_b+\sum_cm_c$ is the level of the state.
Note that if the state is twist even ($N=\mbox{even}$), scalar ($i=\mbox{even}$) and ghost number one ($j=k, (j+k)^2/2=\mbox{even}$), then always $\varepsilon_s=1$, which is used in the analysis of tachyon condensation in level truncation. 



\begin{thebibliography}{999}
\bibitem{Siegel:1984wx}
  W.~Siegel,
  ``Covariantly second-quantized string II,''
  Phys.\ Lett.\ B {\bf 149} (1984) 157
  [Phys.\ Lett.\ B {\bf 151} (1985) 391].

\bibitem{Siegel:1985tw}
  W.~Siegel and B.~Zwiebach,
  ``Gauge String Fields,''
  Nucl.\ Phys.\ B {\bf 263} (1986) 105.

\bibitem{Banks:1985ff}
T.~Banks and M.~E.~Peskin,
``Gauge invariance of string fields,''
Nucl.\ Phys.\ B {\bf 264} (1986) 513.

\bibitem{Itoh:1985bb}
  K.~Itoh, T.~Kugo, H.~Kunitomo and H.~Ooguri,
  ``Gauge Invariant Local Action of String Field from BRS Formalism,''
  Prog.\ Theor.\ Phys.\  {\bf 75} (1986) 162.

\bibitem{Witten:1985cc}
  E.~Witten,
  ``Non-commutative geometry and string field theory,''
  Nucl.\ Phys.\ B {\bf 268} (1986) 253.

\bibitem{Hata:1986zq}
  H.~Hata, K.~Itoh, T.~Kugo, H.~Kunitomo and K.~Ogawa,
  ``Gauge invariant action of interacting string field,''
  Nucl.\ Phys.\ B {\bf 283} (1987) 433.

\bibitem{Batalin:1984jr}
  I.~A.~Batalin and G.~A.~Vilkovisky,
  ``Quantization of gauge theories with linearly dependent generators,''
  Phys.\ Rev.\ D {\bf 28} (1983) 2567
  [Erratum-ibid.\ D {\bf 30} (1984) 508].

\bibitem{Taylor:2003gn}
  W.~Taylor and B.~Zwiebach,
 ``D-branes, tachyons, and string field theory,''
  arXiv:hep-th/0311017.

\bibitem{Schnabl:2005gv}
  M.~Schnabl,
 ``Analytic solution for tachyon condensation in open string field theory,''
  arXiv:hep-th/0511286.

\bibitem{Kato:1983im}
M.~Kato and K.~Ogawa,
``Covariant quantization of string based on BRS invariance,''
Nucl.\ Phys.\ B {\bf 212}, 443 (1983).

\bibitem{AK2}
M.~Asano and M.~Kato,
``Level truncated tachyon potential in various gauges,''
preprint UT-Komaba/06-13 to appear.


\bibitem{Bochicchio:1986zj}
  M.~Bochicchio,
  ``Gauge fixing for the field theory of the bosonic string,''
  Phys.\ Lett.\ B {\bf 193} (1987) 31.

\bibitem{Bochicchio:1986bd}
  M.~Bochicchio,
  ``String field theory in the Siegel gauge,''
  Phys.\ Lett.\ B {\bf 188} (1987) 330.


\bibitem{Thorn:1986qj}
  C.~B.~Thorn,
  ``Perturbation theory for quantized string fields,''
  Nucl.\ Phys.\ B {\bf 287}, 61 (1987).





\bibitem{Asano:2005fm}
  M.~Asano, M.~Kato and M.~Natsuume,
  ``Physical state representations and gauge fixing in string theory,''
  JHEP {\bf 0511} (2005) 033
  [arXiv:hep-th/0509188].


\end{thebibliography}
\end{document}